\documentclass[a4paper,10pt,onecolumn,divps]{article}
\usepackage{algorithm,algorithmic}

\title{Embedding Secret Data in Html Web Page}

\author{Sandipan Dey,
\\Cogniant Technology Solutions
\\\emph{sandipan.dey@gmail.com}
\\
\and Hameed Al-Qaheri\footnote{Contact Author}
\\Department of Quantitative Methods and Information Systems
\\College of Business Administration
\\Kuwait University
\\\emph{alqaheri@cba.edu.kw}
\\
\and Sugata Sanyal,
\\School of Technology and Computer Science
\\Tata Institute of Fundamental Research,
\\Homi Bhabha Road, Mumbai - 400005, India
\\\emph{sanyal@tifr.res.in}
\\}

\date{}

\usepackage{graphicx}

\begin{document}

\maketitle

\section*{Abstract}
\par In this paper, we suggest a novel data hiding technique in an Html Web page. Html Tags are
case insensitive and hence an alphabet in lowercase and one in uppercase present inside an html tag
are interpreted in the same manner by the browser, i.e., change in case in an web page is imperceptible
to the browser. We basically exploit this redundancy and use it to embed secret data inside an web page,
with no changes visible to the user of the web page, so that he can not even suspect about the data hiding.
The embedded data can be recovered by viewing the source of the html page. This technique can easily be
extended to embed secret message inside any piece of source-code where the standard interpreter of that
language is case-insensitive.

\section{Introduction}
\par
Some techniques for hiding data in executables are already proposed (e.g., Shin et al \cite{r4}). In this paper we introduce
a very simple technique to hide secret message bits inside source codes as well. We describe our steganographic technique
by hiding inside html source as cover text, but this can be easily extended to any case-insensitive language source codes.
Html Tags are basically directives to the browser and they carry information regarding how to structure and
display the data on a web page. They are not case sensitive, so tags in either case (or mixed case) are interpreted
by the browser in the same manner (e.g., ``$<head>$'' and ``$<HEAD>$'' refers to the same thing). Hence, there is a redundancy and
we can exploit this redundancy. To embed secret message bits into html, if the cases of the tag alphabets in html cover text
are accordingly manipulated, then this tampering of the cover text will be ignored by the browser and hence it will be imperceptible
to the user, since there will not be any visible difference in the web page, hence there will not be any suspect for it as well.
Also, when the web page is displayed in the browser, only the text contents are displayed, not the tags (those can only
be seen when the user does `view source'). Hence, the secret messages will be kind of hidden to user.

\par
Both redundancy and imperceptibility conditions for data hiding are met, we use these to embed data in html text.
If we do not tamper the html text data that is to be displayed by the browser as web page (this html cover text is analogical to
the cover image, when thought in terms of steganographic techniques in images \cite{r1, r2, r3}),
the user will not even suspect about hidden data in text. We shall only change the case of every character within these Html tags (elements)
in accordance with the secret message bits that we want to embed inside the html web page. If we think of the browser interpreter as a function,
$f_B : \Sigma^{*} \rightarrow \Sigma^{*}$ we see that it is non-injective, i.e., not one to one, since $f_B(x)=f_B(y)$ whenever
$x \in \{\mbox{`A'} \ldots \mbox{`Z'}\}$, $y \in \{\mbox{`a'} \ldots \mbox{`z'}\}$ and $Uppercase(y)=x$. The extraction process
of the embedded message will also be very simple, one needs to just do `view source' and observe the case-patterns of the text within tags and
can readily extract the secret message (and see the unseen), while the others will not know anything.

\par
The length (in bits) of the secret message to be embedded will be upper-limited by the sum of size of text
inside html tags (here we don't consider attribute values for data embedding. In case we consider attribute values for
data embedding, we need to be more careful, since for some tags we should think of case-sensitivity,
e.g. $<$A HREF=``link.html''$>$, since link file name may be case-sensitive on some systems, whereas, attributes such as
$<$h2 align=``center''$>$ is safe). If less numbers of bits to be embedded, we can embed the information inside Header Tag
specifying the length of embedded data (e.g. `$<$Header $25>$' if the length of secret data to be embedded is $25$ bits)
that will not be shown in the browser (optionally we can encrypt this integer value with some private key).
In order to guarantee robustness of this very simple algorithm one may use some simple encryption on the data to be embedded.

\section{The Algorithm for Embedding}
The algorithm for embedding the secret message inside the html cover text is very simple and straight-forward. First, we need to separate out
the characters from the cover text that will be candidates for embedding, these are the case-insensitive text characters inside Html tags.
Figure 2 shows a very simplified automata for this purpose.

\par We define the following functions before describing the algorithm:
\begin{itemize}
\item $l:\Sigma^{*} \rightarrow \Sigma^{*}$ is defined by,
\par
$l(c) = \left\{
\begin{array}{c l}
  ToLower(c) & c \in \{\mbox{`A'}..\mbox{`Z'} \} \\
  c & otherwise
\end{array}
\right\}$
where $ToLower(c) = c + 32$

\item Similarly, $u:\Sigma^{*} \rightarrow \Sigma^{*}$ is defined by,
\par
$u(c) = \left\{
\begin{array}{c l}
  ToUpper(c) & c \in \{\mbox{`a'}..\mbox{`z'}\} \\
  c & otherwise
\end{array}
\right\}$
where $ToUpper(c) = c - 32$
\end{itemize}

\par Here the ascii value of `A' is $65$ and that of `a' is $97$, with  a difference of $32$.

\par It's easy to see that if the domain $\Sigma^{*}=\{\mbox{`a'}..\mbox{`z'}\} \cup \{\mbox{`A'}..\mbox{`Z'}\}$, then \\
$l:\{\mbox{`A'}..\mbox{`Z'}\} \rightarrow \{\mbox{`a'}..\mbox{`z'}\}$ and $u:\{\mbox{`a'}..\mbox{`z'}\} \rightarrow \{\mbox{`A'}..\mbox{`Z'}\}$,
implies that $l(.) = \overline{u(.)}=\Sigma^{*}-l(.)$.	      	

\par Now, we want to embed secret data bits $b_{1}b_{2}..b_{k}$ inside the case-insensitive text inside the Html Tags. If $c_1 c_2 \ldots c_n$
denotes the sequence of characters inside the html tags in cover text (input html). A character $c_i$ is a candidate for hiding a secret message bit iff
it is an alphabet. If we want to hide the $j^{th}$ secret message bit $b_j$ inside the cover text character $c_i$, the corresponding stego-text will be
defined by the following function $f_{stego}$:

\par
$\forall{c_{i}} \in  \{\mbox{`a'}..\mbox{`z'}\} \cup \{\mbox{`A'}..\mbox{`Z'}\}$, i.e. if IsAlphabet($c_i$) is true,

$f_{stego}(c_{i}) = \left\{
\begin{array}{c l}
	l(c_{i}) & b_{j} = 0 \\
	u(c_{i}) & b_{j} = 1
\end{array}
\right\}$,

\par Hence, we have the following:
\begin{eqnarray}
	\label{eq:}
    c_{i} \in  \{\mbox{`a'}..\mbox{`z'}\} \cup \{\mbox{`A'}..\mbox{`Z'}\} \Rightarrow f_{stego}(c_{i}) = l(c_{i}).\overline{b_{j}} + u(c_{i}).b_{j}, \; \forall{i}
\end{eqnarray}

\par Number of bits ($k$) of the secret message embedded into the html cover text must also be embedded inside the html (e.g., in Header element). The figure 1 and the algorithm \ref{alg:embed} together explain this embedding algorithm.

\begin{figure}[h]
    \label{fig:f1}
    \centering
        \includegraphics[width=14cm,height=7cm]{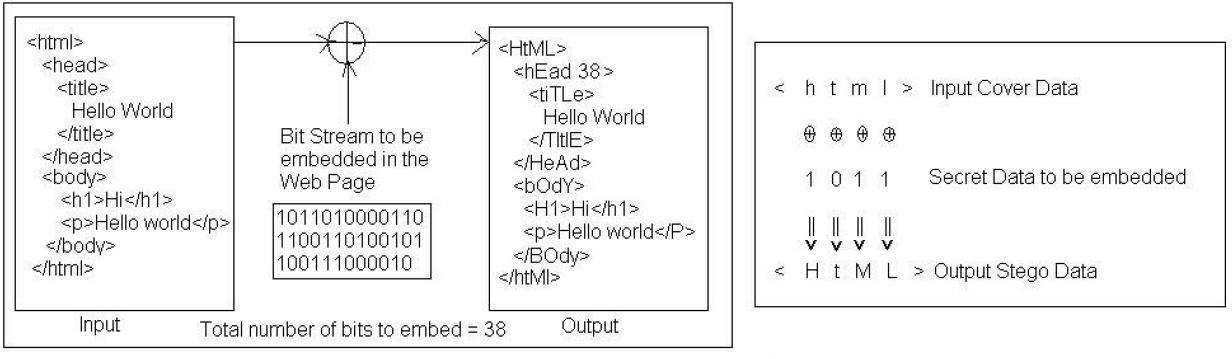}
    \caption{Illustration of how the data hiding works}
\end{figure}

\begin{algorithm} [h]
\begin{algorithmic}[1]
\STATE Search for all the html tags present in the html cover text and extract all the characters $c_1 c_2 \ldots c_n$ from
inside those tags using the DFA described in the figure 2.
\STATE Embed the secret message length $k$ inside html header in the stego text.
\STATE $j \leftarrow 0$.
\FOR {$c_i \in HTMLTAGS, \; i=1 \ldots n$}
\IF {$c_i \in \{\mbox{`a' \dots `z'}\} \cup \{\mbox{`A' \dots `Z'}\}$}
\STATE $f_{stego}(c_{i}) = l(c_{i}).\overline{b_{j}} + u(c_{i}).b_{j}$.
\STATE $j  \leftarrow j + 1$.
\ELSE
\STATE $f_{stego}(c_{i}) = c_{i}$.
\ENDIF
\IF {$j == k$}
\STATE break.
\ENDIF
\ENDFOR
\end{algorithmic}
\caption{Embedding Algorithm} \label{alg:embed} \vspace{-.06 in}
\end{algorithm}

\section{The Algorithm for Extraction}
The algorithm for extraction of the secret message bits will be even more simple. Like embedding process, we must first seperate out the
candidate text (text within tags) that were chosen for embedding secret message bits. Also, we must extract the number of bits
($k$) embedded into this page (e.g., from the header element). One has to use `view source' to find out the stego-text.

\par
Now, we have $d_{i} = f_{stego}(c_{i}), \; \forall{i} \in \{1,2,\ldots,n\}$. If $d_i \in \{\mbox{`a'}..\mbox{`z'}\} \cup \{\mbox{`A'}..\mbox{`Z'}\}$ i.e., an alphabet,
then only it is a candidate for decoding and to extract $b_{i}$ from $d_{i}$, we use the following logic: \\
$b_{i} = \left\{
\begin{array}{c l}
	0 & d_{i} \in \{\mbox{`a' \dots `z'}\} \\
	1 & d_{i} \in \{\mbox{`A' \dots `Z'}\}
\end{array}
\right\}$
\par Repeat the above algorithm $\forall{i} < k$, to extract all the hidden bits.
The figure 2 and the algorithm \ref{alg:extract} together explain this embedding algorithm.

\begin{figure}[h]
    \label{fig:f2}
    \centering
        \includegraphics[width=12cm,height=12cm]{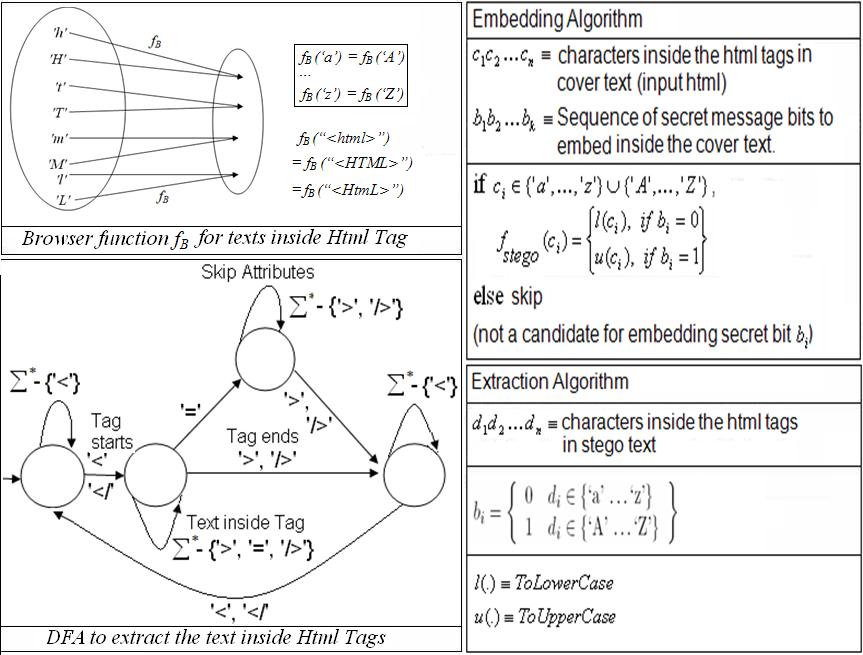}
    \caption{Basic block-diagram for html data-hiding technique}
\end{figure}

\begin{figure}[h]
    \label{fig:htmlsrc-cover}
    \centering
        \includegraphics[width=14cm,height=18cm]{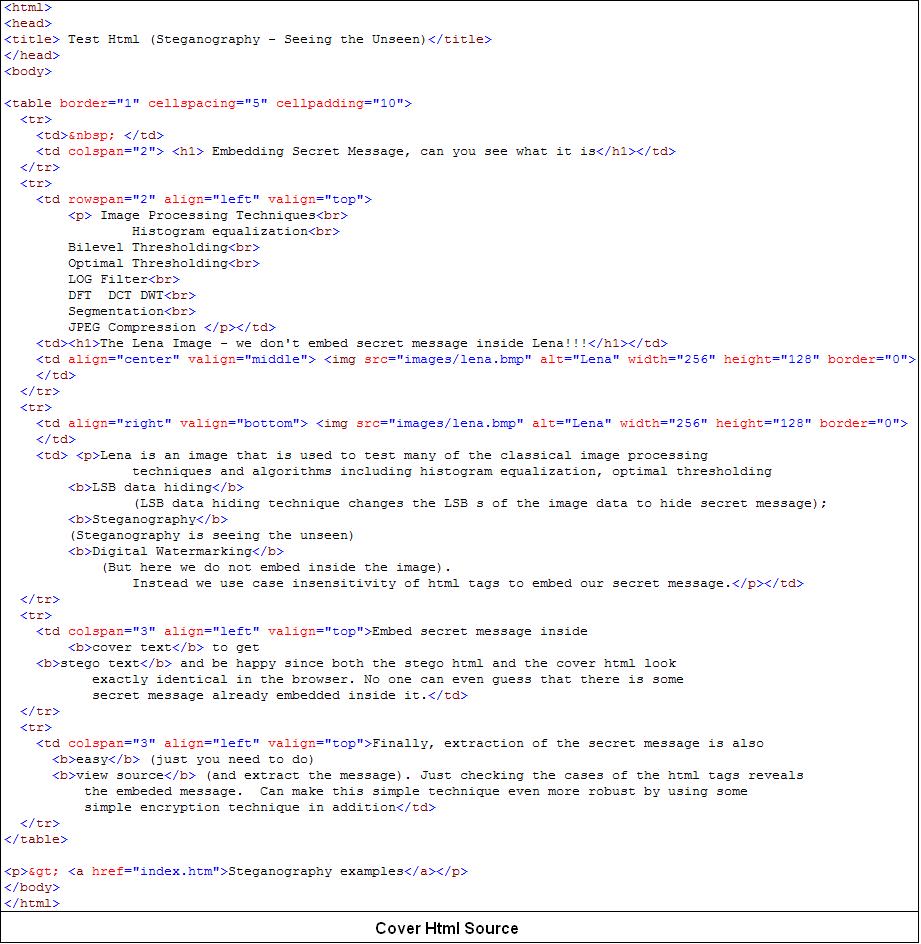}
    \caption{Cover Html source before embedding the secret message}
\end{figure}

\begin{figure}[h]
    \label{fig:htmlsrc-stego}
    \centering
        \includegraphics[width=14cm,height=18cm]{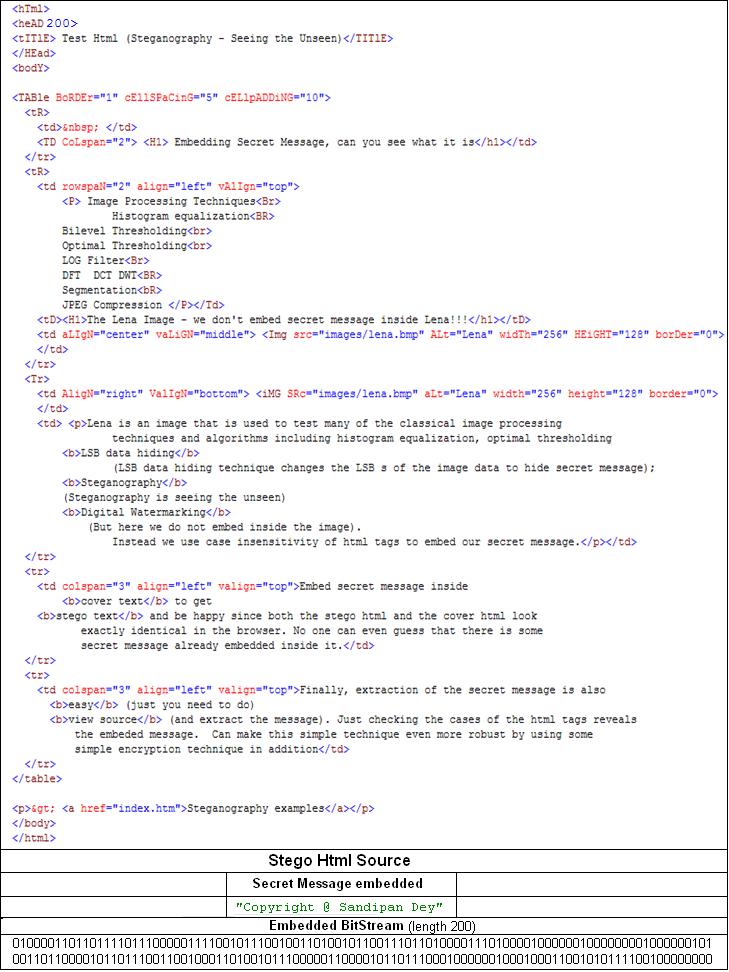}
    \caption{Stego Html source after embedding the secret message}
\end{figure}

\begin{figure}[h]
    \label{fig:html}
    \centering
        \includegraphics[width=14cm,height=18cm]{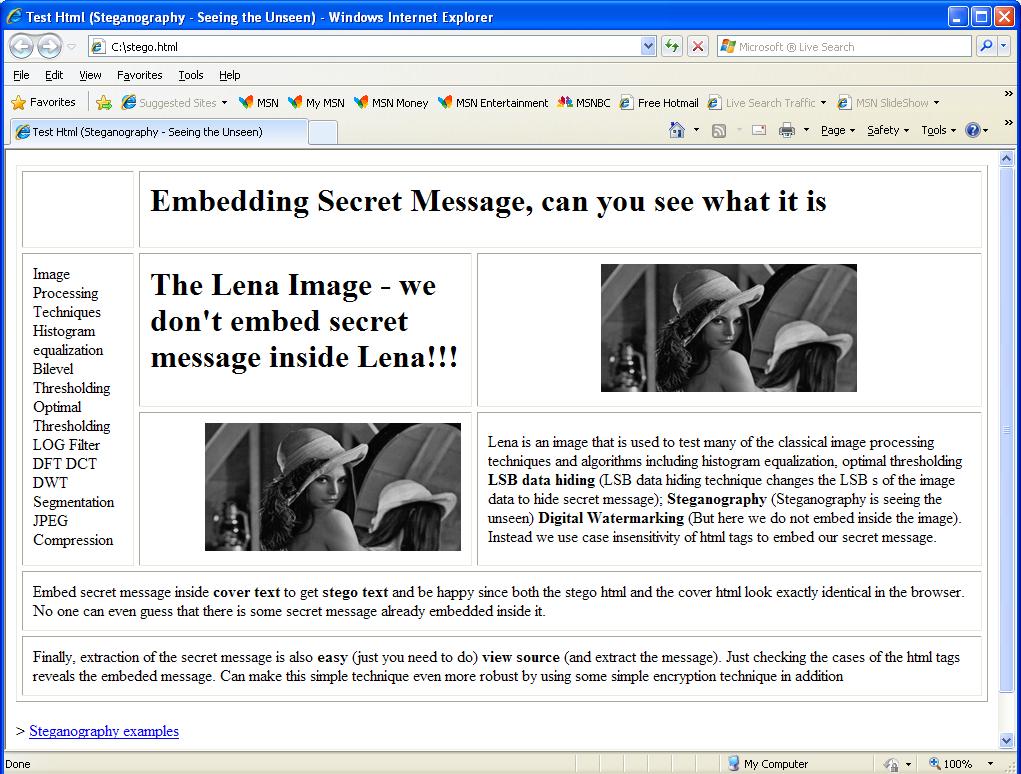}
    \caption{Cover $\&$ Stego Html}
\end{figure}
\clearpage

\begin{algorithm} [h]
\begin{algorithmic}[1]
\STATE Search for all the html tags present in the html stego text and extract all the characters $d_1 d_2 \ldots d_n$ from
inside those tags using the DFA described in the figure 2.
\STATE Extract the secret message length $k$ from inside html header in the stego text.
\STATE $j \leftarrow 0$.
\FOR {$d_i \in HTMLTAGS, \; i=1 \ldots n$}
\IF {$d_i \in \{\mbox{`a' \dots `z'}\}$}
\STATE $b_{j} = 0$.
\STATE $j  \leftarrow j + 1$.
\ELSIF {$d_i \in \{\mbox{`A' \dots `Z'}\}$}
\STATE $b_{j} = 1$.
\STATE $j  \leftarrow j + 1$.
\ENDIF
\IF {$j == k$}
\STATE break.
\ENDIF
\ENDFOR
\end{algorithmic}
\caption{Extraction Algorithm} \label{alg:extract} \vspace{-.06 in}
\end{algorithm}

\begin{figure}[h!]
    \label{fig:graph}
    \centering
        \includegraphics[width=14cm,height=7cm]{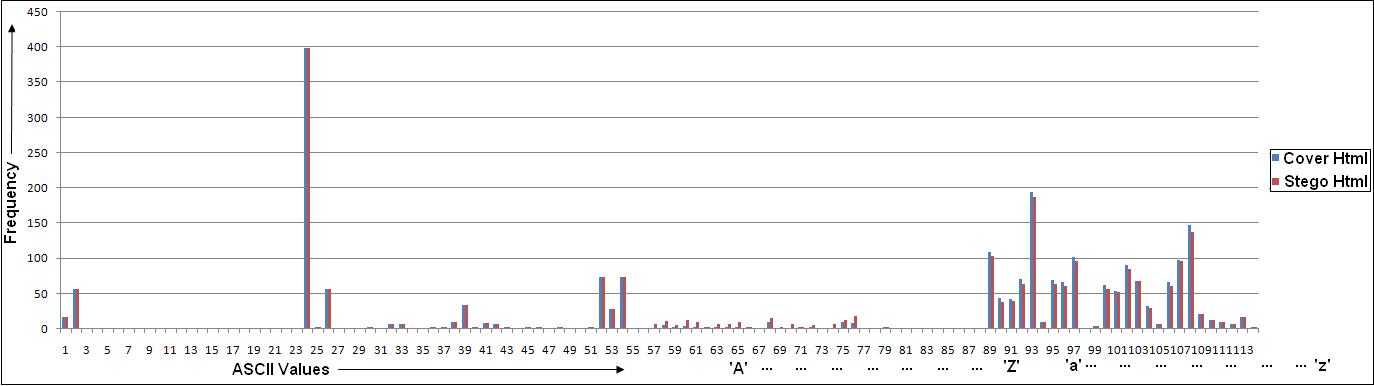}
    \caption{Cover vs Stego Html Histogram}
\end{figure}

\par Figures 3, 4 and 5 show an example of how our method works, while Figure 6 shows the comparison of
the histogram of the cover html and stego html in terms of the (ascii) character frequencies. Classical image hiding techniques
like LSB data hiding technique always introduce some (visible) distortion \cite{r5, r10} in the stego image (that can be reduced using techniques
\cite{r6, r7, r8, r9}), but our data hiding technique in html is novel in the sense that it
introduces no visible distortion in stego text at all.

\section{Conclusions}
In this paper we presented an algorithm for hiding data in html text. This technique can be extended to any
case-insensitive language and data can be embedded in the similar manner, e.g., we can embed secret message
bits even in source codes written in languages like basic or pascal or in the case-insensitive sections (e.g.
comments) in C like case-sensitive languages. Data hiding methods in images results distorted stego-images,
but the html data hiding technique does not create any sort of visible distortion in the stego html text.

\end{document}